\documentclass[english,a4paper]{article}
\usepackage{amsmath}
\usepackage[utf8]{inputenc}
\usepackage{textcomp}
\usepackage{amstext}
\usepackage{graphicx}
\usepackage{setspace}
\usepackage{url}
\usepackage{babel}

\newcommand{\noun}[1]{\textsc{#1}}

\title{Vortexje - An Open-Source Panel Method for Co-Simulation}
\date{March 8, 2013}
\author{
  J. H. Baayen\thanks{Baayen \& Heinz GmbH, Sekr. ER2-1, Hardenbergstra\ss e 36a, 10623 Berlin, Germany, jorn.baayen@baayen-heinz.com.}
}

\begin{document}

\maketitle

\begin{abstract}
This paper discusses the use of the 3-dimensional panel method for dynamical system
simulation. Specifically, the advantages and disadvantages of model
exchange versus co-simulation of the aerodynamics and the dynamical system model are discussed. Based on a trade-off
analysis, a set of recommendations for a panel method implementation
and for a co-simulation environment is proposed. These recommendations
are implemented in a C++ library, offered on-line under an open source
license. This code is validated against XFLR5, and its suitability
for co-simulation is demonstrated with an example of a tethered
wing, i.e, a kite.  The panel method implementation and the co-simulation
environment are shown to be able to solve this stiff problem in a stable fashion.
\end{abstract}

\section{Introduction}

This paper focuses on the application of the 3-dimensional panel method \cite{Hess1967,Maskew1987,Cebeci1999,Katz2001}
to simulation of dynamical systems interacting with fluid flow. The
primary advantage of such a potential flow method over numerical solution of the Navier-Stokes
equations is the ease by which the panel equations can be solved numerically. A panel
method has unknowns on the surface of the immersed body only, whereas numerical
solution of the Navier-Stokes equations requires a 3-dimensional mesh
throughout the region of flow. This reduced computational requirement
renders the panel method a viable candidate for dynamic simulation.

On the other hand, the standard panel method is restricted to the modelling of inviscid, incompressible, and irrotational flows.
This implies that the unmodified panel method is only applicable to subsonic flows
at angles of attack where flow separation does not occur.  Even under these circumstances, the standard
panel method does not predict viscuous drag.  Nevertheless, the panel method has
turned out to be a valuable tool for aerodynamic analysis \cite{Cebeci1999,Katz2001}, and
extensions have been developed to incorporate viscosity effects \cite{Cebeci1999}.

Other 3-dimensional potential flow methods, such as the vortex-lattice and lifting-line
methods, share the same assumptions of inviscid, incompressible, and irrotational flow \cite{Katz2001}.  
These methods, however, introduce the further assumption of an infinitely thin airfoil.  
For this reason we will not consider these methods further, and focus our attention on the 3-dimensional panel method.

Our aim lies in the development of an open-source implementation of
the panel method specifically designed for dynamical system simulation, where
the aerodynamic forces and moments interact with the system dynamics.  An example of such a system is a wing tethered to the ground, i.e.,
a kite. We use the panel method to obtain the aerodynamic forces and
moments acting on the kite, and an ODE solver to compute the shape
and tension of the tether. The aerodynamic forces of the kite act
on the tether, and the other way around. We use this stiff interaction
as a test problem for the development of a panel method and dynamical system
simulation methodology.

Our approach based on the panel method stands in contrast to the approach taken
in \cite{BreukelsPhd}, where a kite is discretized into multiple bodies.  In the
multi-body approach, the aerodynamics for each body are independent, and prescribed
by the modeller.  When the panel method is used, the computed aerodynamic forces and moments
result from the shape of the entire wing, its orientation in the flow, and its interaction
with the wake.  Results obtained from the panel method can therefore be expected to be more accurate.

In the next section, we start by outlining the basic ideas underlying
the panel method.

\section{The Panel Method}

Let a body be immersed in a flow with velocity field $\vec{v}$.  The panel method
models flows that are irrotational.  Irrotationality means that, outside of the body and inside of the flow,
\[
\nabla \times \vec{v} = \vec{0},
\]
or, equivalently \cite{Rudin1964},
\begin{equation}\label{eq:potential-gradient}
\vec{v} = \nabla \phi,
\end{equation}
for some potential function $\phi$. 

In our application, we furthermore assume
the flow to be incompressible, as is applicable for subsonic flows \cite{Katz2001}:
\begin{equation}\label{eq:incompressibility}
\nabla \cdot \vec{v} = 0.
\end{equation}
Substituting Equation \ref{eq:potential-gradient} into Equation \ref{eq:incompressibility}, we find that the potential function satisfies the Laplace
equation:
\begin{equation}\label{eq:laplace}
\nabla^2 \phi = 0.
\end{equation}
The panel method arrives at a solution of the Laplace equation by summing up certain
elementary solutions located on the body boundary.  In our implementation, we use the
elementary solutions known as the \emph{source} and the \emph{doublet} \cite{Katz2001}.  

Taking the gradient of the potential, we obtain the velocity field.  From the velocity field, in turn, an application of the Bernoulli equation
yields the pressure distribution.

The source and doublet elementary solutions are weighted such that, firstly, on the boundary there results no flow velocity normal to it, and secondly, 
that the resulting potential is zero on the body interior.  To make this proceduce suitable for a digital computer, we discretize the body and its wake into a number of panels.  We then assign each panel
both a source as well as a doublet weight variable, labeled $\sigma$ and $\mu$, respectively.  This results in the so-called source and doublet distributions.  The weight assignment conditions, finally, translate into an asymmetric system of linear equations.   For further details, we refer the reader to \cite{Katz2001}.

\section{Existing Open-Source Panel Method Implementations}

The \noun{XFLR5} package \cite{XFLRweb} is the only available open-source implementation of the 3-dimensional source-doublet
panel method that the author is aware of.  Unfortunately, \noun{XFLR5} features a tight integration
between the method implementation and the user interface.  The user interface is designed for the analysis of fixed flight
conditions.  To use the underlying method implementation for dynamic simulation, it would need to be separated from the rest
of the code base.  This would be an unpractical and time-consuming process.

Instead of going through such a process of scavenging bits and pieces from the \noun{XFLR5} code base, we choose to start from a clean slate. In this way we ensure that the code follows our design goals, rather than the other way around. Furthermore, we aim to design the implementation
such that a graphical user interface could readily be built using our implementation
as a reusable component.  This stands in contrast to \noun{XFLR5}, where the user interface is an integral and hard-to-separate part of the package.

\section{Use Case: A Tethered Wing}

During the last two decades, kites, i.e. tethered wings, have gained
popularity not only for sports, but also for the towing of ships \cite{Naaijen2006},
and for the generation of wind energy on land \cite{Loyd1980,Canale2007}.
In our case we investigate the kite as it poses a stiff simulation
problem, and therefore a challenge for our simulation paradigm.  We will use a kite
simulation to validate our design decisions.

\begin{figure}[h]
\centering
\includegraphics[scale=0.25]{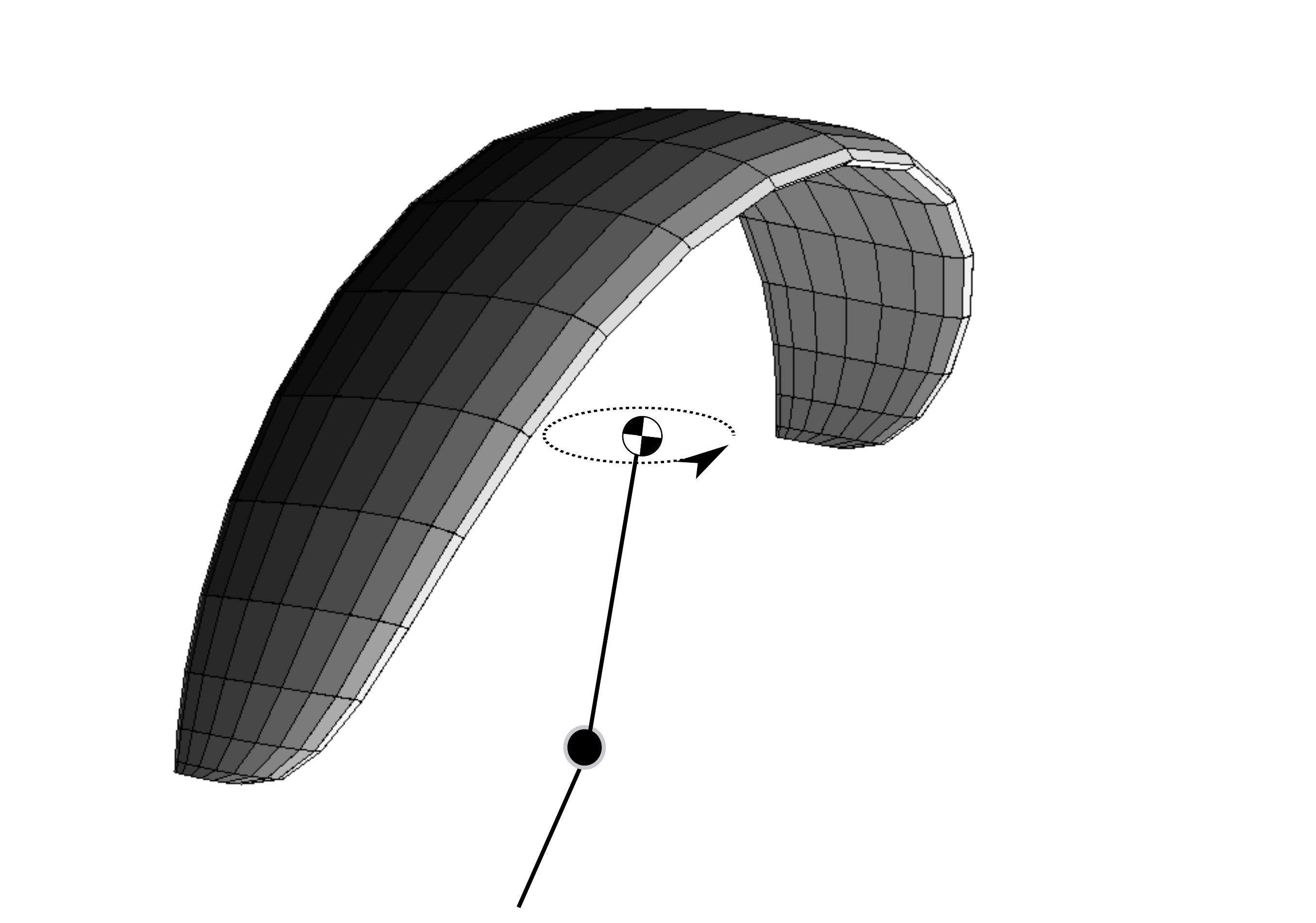}
\caption{Kite and tether simulation.  The yawing direction of the kite is indicated with the dotted arrow.}
\label{fig:kite-mesh}
\end{figure}

We model the kite aerodynamics using the panel method, and the tether
as a string of point-masses connected by high-modulus spring-dampers,
represented by an ODE. The aerodynamic forces acting on the kite,
resulting from the panel method, are added to the spring-damper forces
of the top point-mass in the tether model -- see Figure \ref{fig:kite-mesh}.
In this way the bulk of the kite's lift force is balanced by the line
forces. As a consequence, the aerodynamic forces and the kinematics
operate on different orders of magnitude, resulting in a stiff interaction.

We generated a mesh for a typical C-shaped surf kite with the geometry
specified in Table \ref{table:kite-parameters}. The resulting mesh
is shown in Figure \ref{fig:kite-mesh}. 

\begin{table}[h]
\centering
\begin{tabular}{|l|l|}
\hline 
Parameter & Value\tabularnewline
\hline 
\hline 
Airfoil & Clark-Y\tabularnewline
\hline 
Aspect ratio & $6.0$\tabularnewline
\hline 
Projected aspect ratio & $4.5$\tabularnewline
\hline 
Tip alignment & Trailing edge\tabularnewline
\hline 
Root chord & $3$ m\tabularnewline
\hline 
Tip/chord ratio & $0.25$\tabularnewline
\hline 
Airfoil panels & $18$\tabularnewline
\hline 
Spanwise panels & $18$\tabularnewline
\hline
Tether attachment point & $0.75$ m from leading edge \tabularnewline
\hline 
Mass & $6.0$ kg\tabularnewline
\hline
Yawing inertia & $5.0$ kg $\cdot$ m$^2$ \tabularnewline
\hline
\end{tabular}\medskip{}
\caption{Kite parameters}
\label{table:kite-parameters}
\end{table}

Such surf kites yaw around their tether (cf. Figure \ref{fig:kite-mesh}) in response to asymmetrical deformations \cite{BreukelsPhd}.
To obtain such asymmetrical states, we set up the following dynamic
deformation in the spanwise coordinate $z_b$:
\[
\Delta z_b=ux_b,
\]
where $x_b$ is a dimensionless chordwise coordinate between 0 and 1.  This simplified deformation is inspired by the observation that the tips of arc-shaped kites bend inward when pulling on their respective steering lines.  In order to focus on the essentials, we neglect any further deformation in the normal direction.

\begin{figure}
\centering
\includegraphics[scale=0.3]{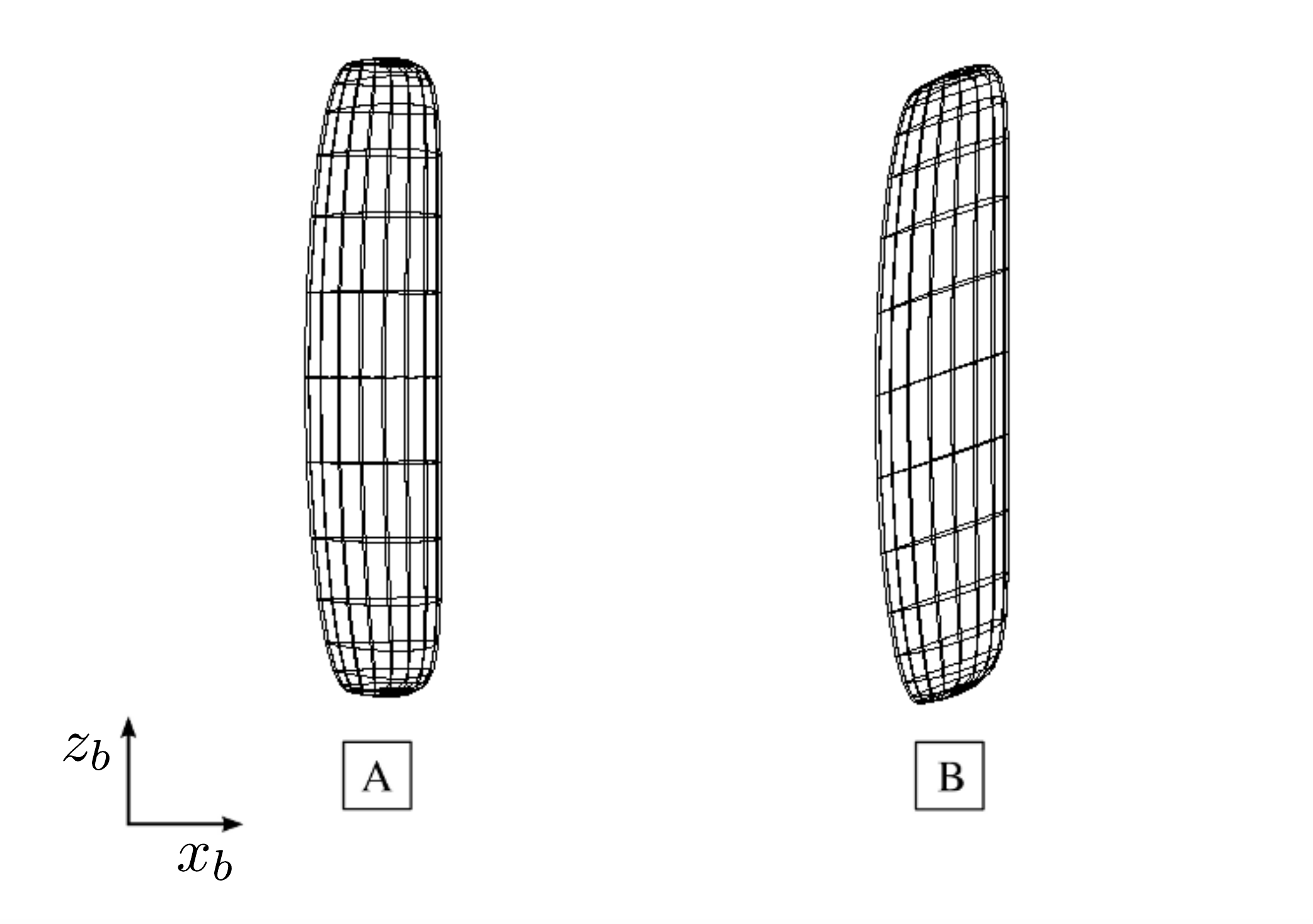}
\caption{Kite mesh without steering deformation (A) and with steering deformation
$u=1$ (B). Top view.}
\label{fig:kite-mesh-steering}
\end{figure}

The steering input $u$ is commanded as a single scalar input variable. The top view of an undeformed mesh, and of a deformed mesh with a
steering input of $u=1$, is displayed in Figure \ref{fig:kite-mesh-steering}. 

Further deformation as well as viscuous effects are neglected. While
viscuous drag is crucial in obtaining an accurate model, in this work
we are primarily interested in obtaining a rudimentary model suitable
for testing our simulation setup.  Empirically identified viscuous drag may be
added to the equations of motion in the dynamical system simulation.

The equations of the tether model are trivial but space-consuming.
For this reason, we do not include them in the present paper.  The
full equations can be obtained from the author on request.

\section{Model Exchange versus Co-Simulation}

In the present section we discuss two different practices that can
be used to integrate the panel method with a dynamical system simulator.
Of primary importance is the distinction between the so-called model exchange
and co-simulation paradigms \cite{Blochwitz2012}.

Model exchange refers to the practice of wrapping an external model
in a stateless function, evaluated as part of an ODE or DAE. The external
model thereby becomes a part of a containing dynamical system model.
In our case, any states such as the source and doublet distributions,
would become part of the system of equations. We note that these states
must indeed be stored, as a proper use of the unsteady Bernoulli equation
does require knowledge of the time-variation of the source and doublet
distributions. The disadvantage of model exchange for the panel method
is that the size of the system matrix would grow quadratically in the number
of panels. Furthermore, all but the most trivial solvers would make
repeated calls to the panel method function. This would pose a performance
problem due to the overhead of solving the -- usually large -- linear
system for computing the doublet distribution.

A co-simulation master, on the other hand, manages two different simulation environments, each
with their own internal states. The co-simulation master coordinates the interaction
between the simulation environments at pre-defined events or times.  The simulators
are commanded to step forward in turns \cite{Blochwitz2012}.  In a co-simulation environment,
a panel method simulator is able to maintain internal states without expanding the system matrices of
the dynamical system simulator.

Stiff interactions between the panel method and the dynamical system
simulator pose a challange for both concepts. In a model exchange
environment, performance of the overall system solver would be limited
by an implicit solver evaluating the panel method Jacobian using finite
differences. The same holds for co-simulation context, where in this
case a specialized implicit co-simulation solver would need to be
developed in addition.

Comparing our two options we see that both suffer from the panel method's
linear system overhead equally. Co-simulation, however, does not suffer
from quadratic growth of the system matrix. This advantage compels
us to focus on the co-simulation paradigm.

\section{Panel Method API Design}

As our focus lies with dynamic simulation, rather than with highly
detailed aerodynamic analysis, we implement a classical first-order
source-doublet panel method as described in \cite{Loyd1980}. For
computation of the surface velocities we follow the treatment of N. Marcov \cite{Dragos2010},
and for 3-D mesh generation we follow the algorithm detailed in \cite{Cebeci1999}. 

Furthermore, during a dynamic simulation a body may find itself interacting
with its own wake. Therefore we supplement the classical method
with singularity elimination following Dixon \cite{Dixon2008}. 

For now, we will not consider aeroelastacity and boundary layer models.  We may, however, add such features in the future.

\begin{figure}[h]
\centering
\includegraphics[scale=0.6]{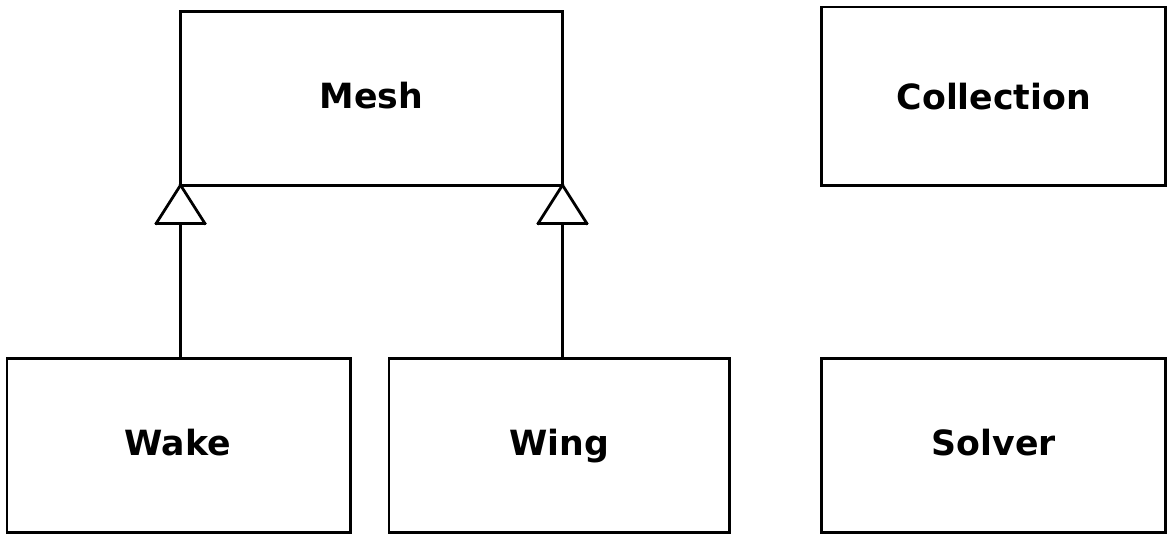}
\caption{Class diagram.}
\label{fig:vortexje-uml}
\end{figure}

We will now briefly discuss the considerations leading to our software
architecture. In order to facilitate a modular description of an object
in flow, we choose to allow for a multitude of meshes to be specified.
Wake-emitting, wake, and source-only meshes are allowed. In line with
the common approach of modeling an airplane as consisting of wake-emitting
and source-only parts \cite{Cebeci1999,Katz2001}, we choose to implement
an additional class for collecting a multitude of meshes. In this
way, the programmer may handle the airplane as one, applying transformations
to the collection as if it were a single object.  These design considerations are shown in the diagram shown in Figure
\ref{fig:vortexje-uml}.

The classes were implemented in C++ in a library called \noun{Vortexje},
available on-line \cite{Vortexjeweb} under the terms of an open-source
license.  Its sole dependency, aside from a C++ compiler, is the \noun{Eigen} 
\cite{Eigenweb} template library for linear algebra.  \noun{Vortexje} uses the
\noun{Bi-CGSTAB} \cite{vanDerVorst1992} implementation from \noun{Eigen} 
\cite{Eigenweb} to solve the doublet coefficient equations.
An example code using the library is listed on the \noun{Vortexje} website \cite{Vortexjeweb}.

In the next section we compare simulation results from \noun{Vortexje}
and \noun{XFLR5} for a simple example, before, in the section thereafter,
embarking on a co-simulation of a kite with its tether.

\section{Validation}

To validate our implementation, we simulate a simple reference wing using both \noun{XFLR5} (in inviscid mode) 
and \noun{Vortexje}.  Since the use of \noun{XFLR5} is well-established,
we expect it to compute the pressure distribution on our reference wing correctly.  For this reason,
we test the pressure distribution and resulting aerodynamic coefficients computed by \noun{Vortexje}
against those computed by \noun{XFLR5}.  

\begin{figure}[h]
\centering
\includegraphics[scale=0.33]{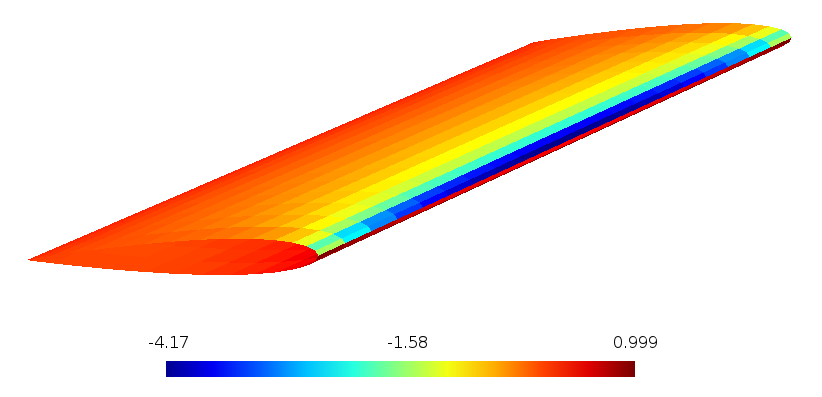}
\caption{Pressure distribution, as computed by \noun{Vortexje}.}
\label{fig:pressure-distribution}
\end{figure}

\begin{figure}[h]
\centering
\includegraphics[scale=0.6]{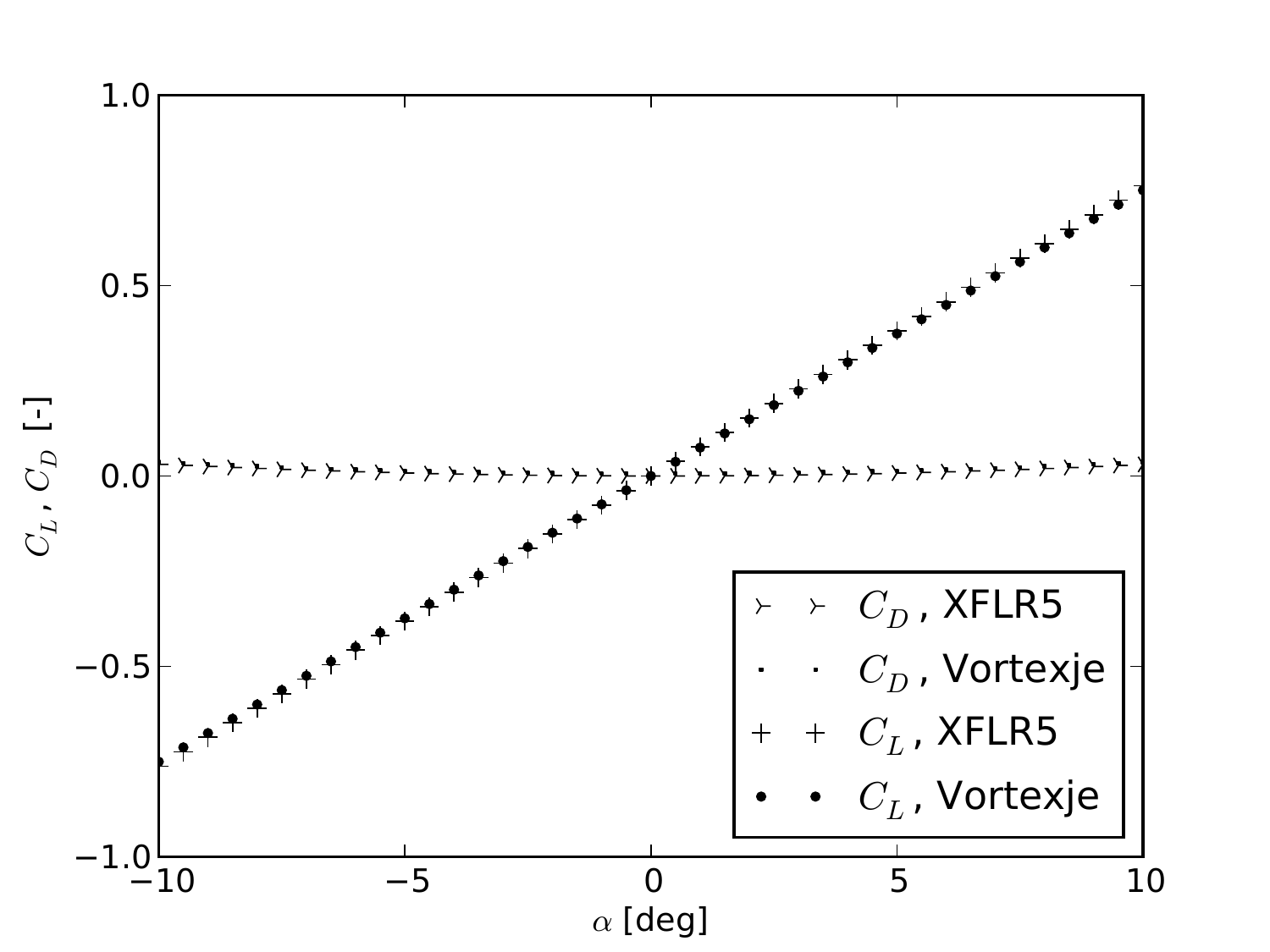}
\caption{Comparison of \noun{XFLR5} and \noun{Vortexje} simulation results.}
\label{fig:validation-results}
\end{figure}

Our reference wing is specified in Table \ref{table:validation-setup}.  The pressure distribution 
of this wing, computed by \noun{Vortexje} at an angle of attack of $10$ degrees,
is shown in Figure \ref{fig:pressure-distribution}.  A comparison of the integrated
pressure distributions of \noun{XFLR5} and \noun{Vortexje} is shown in
Figure \ref{fig:validation-results}. 

\begin{table}
\centering{}%
\begin{tabular}{|l|l|}
\hline 
Parameter & Value\tabularnewline
\hline 
\hline 
Airfoil & NACA0012\tabularnewline
\hline 
Span & $6$ m\tabularnewline
\hline 
Chord & $1$ m\tabularnewline
\hline 
Airfoil panels & $32$\tabularnewline
\hline 
Spanwise panels & $40$\tabularnewline
\hline 
Airspeed & $30$ m/s\tabularnewline
\hline 
Fluid density & $1.2$ kg/m$^{3}$\tabularnewline
\hline 
\end{tabular}\medskip{}
\caption{Validation set-up.}
\label{table:validation-setup}
\end{table}

On closer inspection of the subtle differences shown in Figure \ref{fig:validation-results},
we find that \noun{XFLR5} computes the potential gradient on the
wing surface as the gradient of the doublet distribution:
\[
\vec{v}(\vec{x})=-(\nabla\mu)(\vec{x}).
\]
This assumption originates from \cite{Maskew1987}, where it is stated
without proof. We, on other hand, implemented N. Marcov's formula
\cite{Dragos2010}, which derives an alternative expression for the
surface-derivative of the potential:
\begin{equation}
\vec{v}(\vec{x})=\vec{w}(\vec{x})-\frac{1}{2}(\nabla\mu)(\vec{x}),\label{eq:marcov-surface-velocity}
\end{equation}
with the term $\vec{w}(\vec{x})$ representing the Cauchy principal
value of the potential gradient contribution of the entire boundary.
To compute this value numerically, we sum up the potential gradients
of all panels, treating the local panel the same as all others. We
do not run into the theoretical singularity this way, as the explicit
expressions of the potential gradients for source and doublet panels
are singular only on panel boundaries \cite{Katz2001}. The \emph{theoretical}
singularity of the local potential contribution at the point of evaluation,
however, is significant, and gives rise to the second term in Equation
\ref{eq:marcov-surface-velocity}. 

There is one more difference in implementation.  This pertains the so-called
far-field formulas \cite{Katz2001}.  The far-field formulas are computationally inexpensive approximations of the panel potential contributions for points far away from the respective panel.  Our omission of the far-field formulas is likely to share responsiblity for the subtle divergence in results of Figure \ref{fig:validation-results}.

Based on the fact that the observed differences are extremely small, we consider 
\noun{Vortexje} to have passed our validation test for the reference wing.  Furthermore,
our implementation rests on a mathematically underpinned surface-derivative formula, and its results may
therefore be more reliable than those provided by \noun{XFLR5}.

\section{Co-Simulation Methodology}

We now return to co-simulation, and verify our approach with
our case problem, the tethered kite. 

For encoding the tether ODE, we selected \noun{Modelica} for its ease
of use and extensibility. As the bulk of the computational time will
be spent in the ODE and panel method solvers, we choose to implement
the co-simulation master in the \noun{Python} scripting language.
The \noun{JModelica} project \cite{Aakesson2009} maintains an open-source
Python package for simulating model exchange Functional Mock-up Units (FMUs), \noun{PyFMI} \cite{Blochwitz2012}.

We also use \noun{JModelica} to compile the tether model to an FMU,
and use \noun{PyFMI} to access the FMU from Python. \noun{JModelica}'s
\noun{Assimulo} package -- which wraps the \noun{SUNDIALS} suite of
integrators \cite{Hindmarsh2005} -- is used to integrate the tether
model.

To the upper end of the tether, we attach a kite mesh with predefined
deformation states in \noun{C++}.  The presence of multiple lines is emulated
by prescribing zero roll and pitch angles with respect to the tether.  The yaw angle, 
on the other hand, is determined by the aerodynamic moments acting on the kite mesh.  See Figure \ref{fig:kite-mesh}.

\begin{figure}[h]
\centering
\includegraphics[scale=0.2]{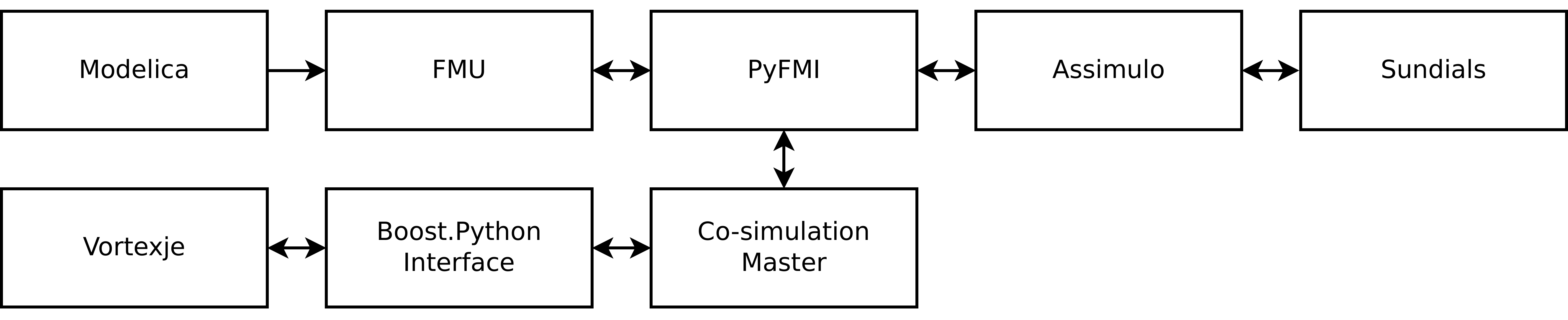}
\caption{Co-simulation architecture.}
\label{fig:cosimulation-arch}
\end{figure}

The interaction between kite and tether might, in theory, be handled using a Functional Mock-up Interface (FMI) co-simulator\footnote{The Functional Mock-up Interface (FMI) is not to be confused with Functional Mock-up Units (FMUs).  FMUs implement the FMI. }.  As of this writing, however, \noun{PyFMI}
does not support the FMI co-simulation standard.  Instead, we use
a \noun{Boost.Python \cite{Boostweb}} class to set up communication
between \noun{Python} and \noun{Vortexje}. Figure \ref{fig:cosimulation-arch}
summarizes the complete set-up.

\begin{figure}[h]
\begin{centering}
\includegraphics[scale=0.225]{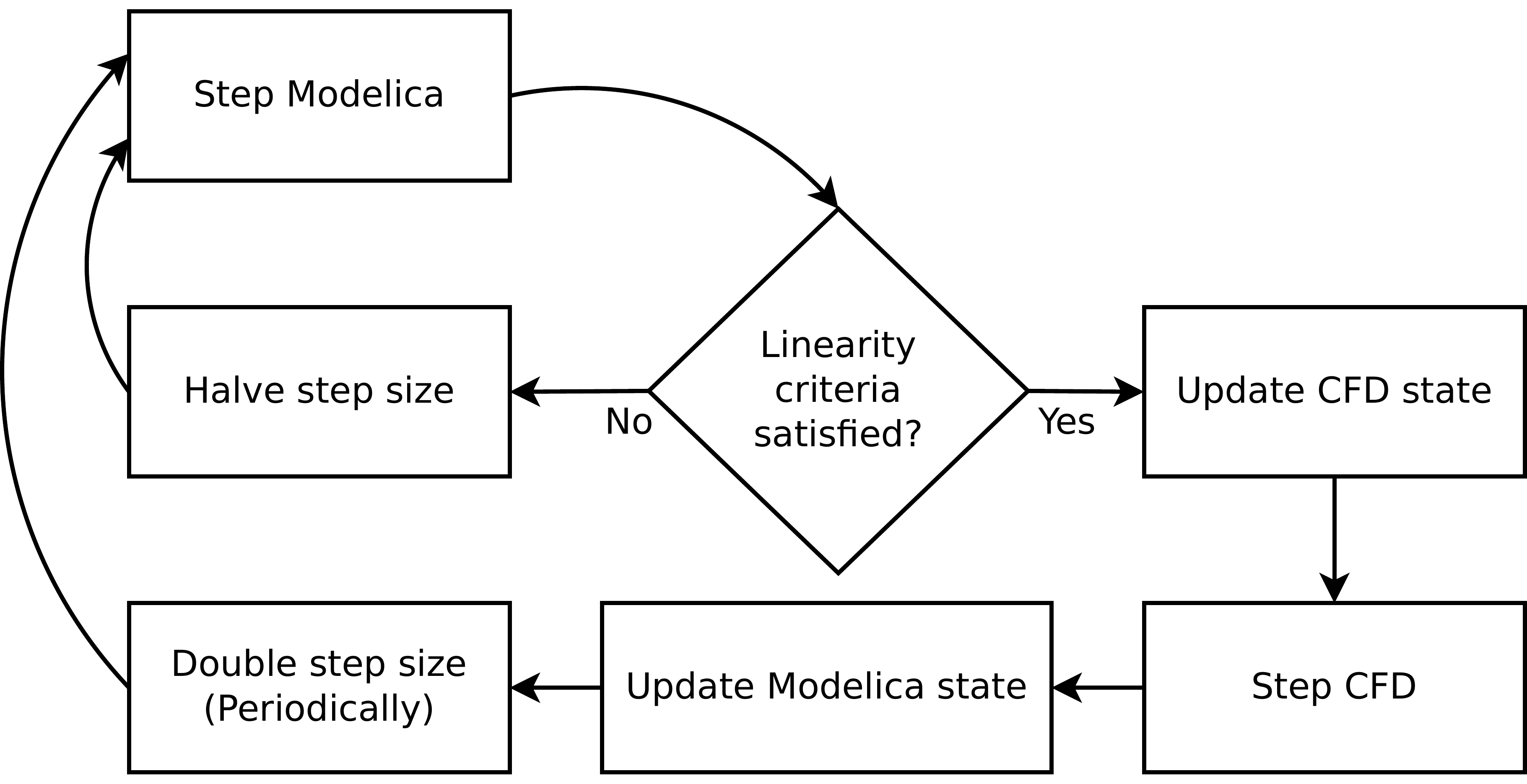}\caption{Co-simulation flow chart.}
\label{fig:cosimulation-flow}
\end{centering}
\end{figure}

Both simulators are stepped in turn. The step size $h$ is reduced on a
binary logarithmic scale until roughly linear behaviour is achieved:
\[
\frac{\left|\|y(t+h)-y(t)\|-\|\dot{y}\|h\right|}{\|\dot{y}\|h}<\varepsilon,
\]
where $y$ denotes the state of the dynamical system, and $\varepsilon$ is a pre-defined linearity tolerance value. The
step size reduction is carried out down to a minimum step size $h_{0}$.
Every $T_{h}$ time steps the step size is doubled, up to a maximum
step size $h_{1}$, before entering the step size reduction algorithm.
In this way, we ensure that the simulator does not remain stuck with
a very small step size after integrating an interval of high stiffness,
while at the same time refraining from trying an excessively large
step size in vain too often. This stepping scheme is summarized as
a flow chart in Figure \ref{fig:cosimulation-flow}. The parameters
values selected for the solver are shown in Table \ref{table:solver-parameters}.

\begin{table}[h]
\centering{}%
\begin{tabular}{|l|l|}
\hline 
Setting & Value\tabularnewline
\hline 
\hline 
Minimum master step size $h_{0}$ & $1\times10^{-6}$\tabularnewline
\hline 
Maximum master step size $h_{1}$ & $5\times10^{-3}$\tabularnewline
\hline 
Step size doubling period $T_{h}$ & $10$\tabularnewline
\hline 
Linearity tolerance $\varepsilon$ & $0.1$\tabularnewline
\hline 
\noun{SUNDIALS} solver & \noun{CVODE}\tabularnewline
\hline 
CVODE absolute tolerance & $1\times10^{-6}$\tabularnewline
\hline 
CVODE relative tolerance & $1\times10^{-6}$\tabularnewline
\hline 
Bi-CGSTAB maximum iterations & $20\times10^{3}$\tabularnewline
\hline 
Bi-CGSTAB tolerance & $1\times10^{-10}$\tabularnewline
\hline
\end{tabular}\medskip{}
\caption{Solver parameters.}
\label{table:solver-parameters}
\end{table}

\begin{figure}[h]
\centering
\includegraphics[scale=0.6]{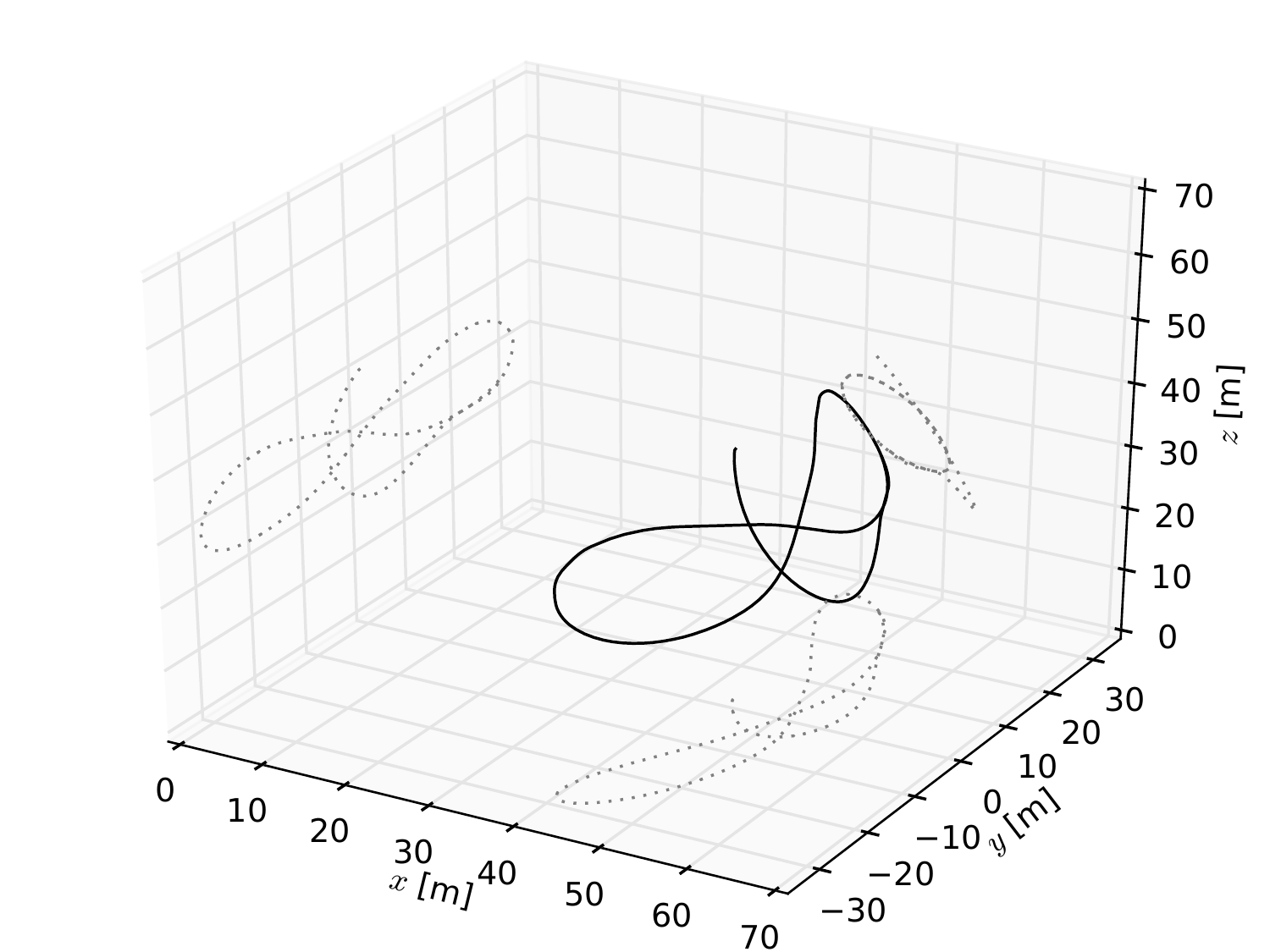}
\caption{Co-simulated kite trajectory.}
\label{fig:kite-trajectory}
\end{figure}

With the control algorithm from \cite{Baayen2012, Baayen2012-3} and an environmental wind velocity of $6$ m/s, the trajectory
of the kite is shown in Figure \ref{fig:kite-trajectory}.  The ``tail'' in said figure is the arc traced from the initial point to the target trajectory.  No numerical instability was observed.

\section{Conclusions and Future Directions}

In this work we proposed a design for a panel method implementation
capable of efficient co-simulation. The design was implemented in
a C++ code, released on-line under an open-source license. We continued
with validation results, followed by an analysis of a co-simulation
of a kite with its tether. This tether is composed of point-masses
and spring-dampers, and the interaction of the spring-damper and aerodynamic
kite forces results in a stiff problem. We showed that our co-simulation
framework -- implemented entirely using open-source software -- is
able to solve the kite problem in a stable fashion. 

Future work will focus on adding optional aeroelasticity and boundary
layer simulation features to the software. We invite interested readers
to evaluate our software, and hope to establish a community of users
and developers.

\bibliographystyle{hplain}
\bibliography{cosimulation}

\end{document}